# Active Cyber Defense Dynamics Exhibiting Rich Phenomena

Ren Zheng
Fudan University
zhrkevin@gmail.com

Wenlian Lu
Fudan University
wenlian@fudan.edu.cn

Shouhuai Xu
UT San Antonio
shxu@cs.utsa.edu

## ABSTRACT

The Internet is a man-made complex system under constant attacks (e.g., Advanced Persistent Threats and malwares). It is therefore important to understand the phenomena that can be induced by the interaction between cyber attacks and cyber defenses. In this paper, we explore the rich phenomena that can be exhibited when the defender employs active defense to combat cyber attacks. To the best of our knowledge, this is the first study that shows that *active cyber defense dynamics* (or more generally, *cybersecurity dynamics*) can exhibit the bifurcation and chaos phenomena. This has profound implications for cyber security measurement and prediction: (i) it is infeasible (or even impossible) to accurately measure and predict cyber security under certain circumstances; (ii) the defender must manipulate the dynamics to avoid such *unmanageable situations* in real-life defense operations.

## 1. INTRODUCTION

Malicious attacks in cyberspace will remain to be a big problem for the many years to come. This is fundamentally caused by the complexity of the Internet and computer systems (e.g., we cannot assure that a large software system has no security vulnerabilities). It is therefore important to understand and characterize the phenomena that can be exhibited at the global level of a cyber system, ranging from an enterprise network to the entire cyberspace. The emerging framework of *Cybersecurity Dynamics* [34, 35, 7, 4] offers a systematic approach for understanding, characterizing, and quantifying the phenomena as well as cyber security in general.

The current generation of cyber defenses is often based on *reactive* tools that are known to have limited success. For example, infected/compromised computers cannot be cleaned up even by using *multiple* anti-malware tools together [21]. Moreover, reactive defense has a fundamental limitation, namely that the effect of attacks is automatically amplified by network connectivity, but the effect of reactive defenses is not. This *attack-defense asymmetry* had been implied by studies such as [29, 6, 3, 28, 39], but was not explicitly pointed out until [36].

One approach to overcoming the aforementioned attack-defense asymmetry is to adopt *active* cyber defense, which is to use the same mechanism that is exploited by attackers. More specifically, active defense aims to spread some "white" worms (called *defense-ware* in this paper) to automatically identify and "kill" the malicious malwares in compromised/infected computers [2, 1, 30, 26, 16, 18, 14, 31]. In some sense, active cyber defense already takes place in cyberspace because (for example) the malware called `Welchia` attempts to "kill" the malware called `Blaster` in compromised computers [26, 22], but it may take some years for full-scale active cyber defenses to arise [18, 27, 32]. The first mathematical model for studying the *global effectiveness* of active cyber defense has been proposed recently [36]. In this paper, we further the study of active cyber defense dynamics from a *new* perspective.

**Our contributions.** We substantially extend some aspects of the first mathematical model of active cyber defense dynamics [36] (to be fair, we should note that [36] offers some perspectives that are not considered in our model as well). The extensions can be characterized as follows. First, we accommodate more general *attack-power* and *defense-power* functions, meaning that our results are applicable to a broader setting than what is investigated in [36]. Second, we allow the *attack network structure* to be different from the *defense network structure*, which are assumed to be identical in [36]. This is important and realistic because the attack-defense interaction structures are often "overlay" networks on top of some physical networks, and as such, the defender and the attacker can use different structures based on their own defense/attack strategies.

The extended model allows us to explore the rich phenomena that can be exhibited by active cyber defense dynamics. Specifically, we show that active cyber defense dynamics can exhibit the bifurcation and chaos phenomena (we call them *unmanageable situations* in cyber security). To the best of our knowledge, this is the first study that shows that bifurcation and chaos are relevant in the cyber security domain. These phenomena indicate limitations on the measurement and prediction of cyber security, and highlight that cyber defenders must manipulate the (active) cyber defense dynamics to avoid such unmanageable situations in real-life cyber defense operations.

**Disclaimer.** The active cyber defense strategy explored in the present paper does not advocate that defenders should retaliate from attackers, because it is well known that the attackers, or more precisely the IP addresses that are launching attacks against the victims, could well be victims that are abused by the real attackers as stepping stones. Moreover, defensewares (i.e., "white" worms) are meant to clean up the compromised computers, not to compromise the secure computers. Most important of all, the active defense operations should be contained within the networks under the defender's jurisdiction (e.g., an enterprise network defender may use active defense to clean up the enterprise network but not going beyond the enterprise's perimeter). This can be assured, for example, by making the enterprise's computers and firewalls recognize defensewares via digital signatures. This means that the enterprise computers will only run defensewares that are accompanied with digital signatures that can be verified by the computers' hardware via an embedded signature verification key, and that the firewall recognizes and blocks out-bound defensewares.

The rest of the paper is organized as follows. In Section 2, we present our active cyber defense dynamics model. In Section 3, we analyze equilibria (or attractors) of active cyber defense dynamics. In Section 4, we explore the transition between attractors. In Section 5, we investigate the emergence of bifurcation. In Section 6, we explore the chaos phenomenon. We discuss related prior work in Section 7 and conclude the paper in Section 8.

The main notations we use are summarized as follows.

| | |
|---|---|
| $\mathbb{R}, \mathbb{R}^+, \mathbb{C}$ | the sets of real numbers, positive real numbers and complex numbers, respectively |
| $\Re(\omega), \Im(\omega)$ | the real and imaginary parts of complex number $\omega \in \mathbb{C}$, respectively |
| $I_n$ | the $n \times n$ identity matrix |
| $G_B, A_B$ | $G_B = (V, E_B)$ is the *defense network structure*, $A_B$ is the adjacency matrix of $G_B$ |
| $G_R, A_R$ | $G_R = (V, E_R)$ is the *attack network structure*, $A_R$ is the adjacency matrix of $G_R$ |
| $N_{v,G'}$ | $N_{v,G'} = \{u \in V' : (u,v) \in E'\}$ is the neighbors of $v$ in graph/network $G' = (V', E')$ |
| $\deg(v, G')$ | $\deg(v) = \lvert N_v \rvert$ is node $v$'s in-degree in graph/network $G' = (V', E')$ |
| $D_{A'}$ | $D_{A'} = [d_{vv}]_{n\times n}$ is a diagonal matrix corresponding to adjacency matrix $A' = [a'_{vu}]_{n\times n}$, where $d_{vv} = \sum_{u=1}^{n} a_{vu}$ is the in-degree of node $v$ in graph $G'$ corresponding to $A'$ |
| $\lambda(M)$ | the set of eigenvalues of matrix $M$ |
| $\lambda_1(M)$ | the eigenvalue of $M$ with the largest real part (or $\lambda_1$ when $M$ is clear from the context) |
| $B_v(t), R_v(t)$ | the probability that node $v \in V$ is in sate blue (i.e., secure) and state red (i.e., compromised) at time $t$, respectively |
| $\langle B_v(t) \rangle$ | the average portion of blue nodes at time $t \geq 0$, namely $\langle B_v(t)\rangle = \frac{1}{\lvert V \rvert}\sum_{v \in V} B_v(t)$ |
| $\mathbf{B}(t), \mathbf{R}(t)$ | $\mathbf{B}(t) = [B_1(t), \ldots, B_n(t)]$, $\mathbf{R}(t) = [R_1(t), \ldots, R_n(t)]$, where $n = \lvert V \rvert$ |
| $B^*$ | the homogeneous equilibrium of $\mathbf{B}(t)$ as $t \to \infty$, namely $B_v(t) = \sigma\ \forall v \in V$ as $t \to \infty$ |
| $f(\cdot), g(\cdot)$ | $f(\cdot) : [0,1] \to \{0\} \cup \mathbb{R}^+$ is the defense-power function, $g(\cdot) : [0,1] \to \{0\} \cup \mathbb{R}^+$ is the attack-power function |
| $\theta_{v,BR}(t)$ | the probability that the state of node $v$ changes from blue to red at time $t$ |
| $\theta_{v,RB}(t)$ | the probability that the state of node $v$ changes from red to blue at time $t$ |

# 2. EXTENDED ACTIVE CYBER DEFENSE DYNAMICS MODEL

**Review of the model in [36].** Suppose attacker and defender interact in a cyber system that consists of a finite node population $V = \{1, 2, \cdots, n\}$, where each node can abstract/represent a computer. At any time $t \geq 0$, a node $v \in V$ is in one of two states: blue, meaning that the node is secure but vulnerable to attacks; red, meaning that the node is compromised. For a given cyber system, the attacker spreads computer malwares (e.g., Advanced Persistent Threats) to compromise computers, while the defender spreads *defensewares* (e.g., "white" worms) to detect and clean up (or "cure") the compromised computers. Suppose both the malwares and the defensewares spread over the same *attack-defense network structure*, namely a finite simple graph $G = (V, E)$, where $V = \{1, 2, \cdots, n\}$ is the vertex set mentioned above, and $E$ is the edge set such that $(u, v) \in E$ means (i) a compromised node $u$ can attack a secure node $v$ and (ii) a secure node $u$ can use active defense to detect and clean up a compromised node $v$.

**Our extension to the model in [36].** Rather than assuming the attacker and defender use the same *attack-defense network structure*, we consider two network structures: the *defense network structure* $G_B = (V, E_B)$ over which defensewares spread, and the *attack network structure* $G_R = (V, E_R)$ over which malwares spread. Both network structures are directed or undirected graphs. Specifically, $(u, v) \in E_B$ means a secure node $u$ can use active defense to "cure" a compromised node $v$, and $(u, v) \in E_R$ means a compromised node $u$ can attack a secure node $v$. We do not make any restrictions on the attack/defense network structures, except that we assume $G_B$ and $G_R$ are simple graphs with no self-edges.[1] (For the purpose of illustrating results, we will use random graphs as concrete examples though.)

Denote by $A_B = [a^B_{vu}]_{n\times n}$ the adjacency matrix of $G_B$ where $a^B_{vu} = 1$ if and only if $(u, v) \in E_B$. Denote by $A_R = [a^R_{vu}]_{n\times n}$ the adjacency matrix of $G_R$ where $a^R_{vu} = 1$ if and only if $(u, v) \in E_R$. Note that the representation accommodates both directed and undirected graphs. Denote by $B_v(t)$ and $R_v(t)$ the probability that node $v \in V$ is in state blue (i.e., secure) and state red (i.e., compromised) at time $t$, respectively.

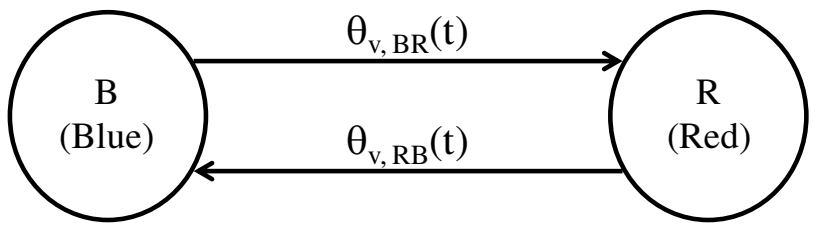

**Figure 1: The state transition diagram for a node $v \in V$.**

Figure 1 depicts the state transition diagram for *individual node* $v \in V$, where $\theta_{v,RB}(t)$ is the probability that node $v$'s state changes from red to blue at time $t$, and $\theta_{v,BR}(t)$ is the probability that node $v$'s state changes from blue to red at time $t$. This leads to the following master equation of active cyber defense dynamics:

$$\begin{cases} \dfrac{dB_v(t)}{dt} = \theta_{v,RB}(t) \cdot R_v(t) - \theta_{v,BR}(t) \cdot B_v(t) \\ \dfrac{dR_v(t)}{dt} = \theta_{v,BR}(t) \cdot B_v(t) - \theta_{v,RB}(t) \cdot R_v(t) \end{cases} \quad (1)$$

In order to specify $\theta_{v,RB}(t)$, we use the concept of *defense-power* function $f(\cdot) : [0, 1] \to \{0\} \cup \mathbb{R}^+$, which abstracts the power of the defenseware in detecting and cleaning up compromised (red) nodes. In order to specify $\theta_{v,BR}(t)$, we use the concept of *attack-power* function $g(\cdot):[0,1]\rightarrow\{0\}\cup\mathbb{R}^+$, which abstracts the power of the malware in compromising secure (blue) nodes. It is intuitive that both defense-power and attack-power functions should be dependent on the defense and attack network structures, respectively. Therefore, we have the following general form:

$$\begin{aligned}\theta_{v,RB}(t) &= f\left(\frac{1}{\deg(v,G_B)}\sum_{u\in N_{v,G_B}} B_u(t)\right),\\ \theta_{v,BR}(t) &= g\left(\frac{1}{\deg(v,G_R)}\sum_{u\in N_{v,G_R}} B_u(t)\right)\end{aligned}$$

where $N_{v,G_B}=\{u:(u,v)\in E_B\}$ is the set of node $v$'s neighbors in graph $G_B$ and $N_{v,G_R}=\{u:(u,v)\in E_R\}$ is the set of node $v$'s neighbors in graph $G_R$.

[1]It is possible to accommodate *privilege escalation* in the present model, by treating a computer as a set of nodes that correspond to different privileges. We leave the details to future investigation.

For the present characterization study, it is sufficient to require that the defense-power and attack-power functions possess some basic properties. First, we have $f(0)=0$ because active defense must be launched from some blue node, and $g(1)=0$ because attack must be launched from some red node. Second, we have $f(x)>0$ for $x\in(0,1]$ because any active defense may succeed, and $g(x)>0$ for $x\in[0,1)$ because any attack may succeed. Third, the two functions do not have to abide by any specific relation, except that they are differentiable (for the sake of analytic treatment).

As a result, the master equation of active cyber defense dynamics, namely Eq. (1), becomes:

$$\begin{aligned}\frac{dB_v(t)}{dt} &= f\left(\frac{1}{\deg(v,G_B)}\sum_{u\in N_{v,G_B}} B_u(t)\right)R_v(t) -\\ &\quad g\left(\frac{1}{\deg(v,G_R)}\sum_{u\in N_{v,G_R}} B_u(t)\right)B_v(t)\\ \frac{dR_v(t)}{dt} &= g\left(\frac{1}{\deg(v,G_R)}\sum_{u\in N_{v,G_R}} B_u(t)\right)B_v(t) -\\ &\quad f\left(\frac{1}{\deg(v,G_B)}\sum_{u\in N_{v,G_B}} B_u(t)\right)R_v(t)\end{aligned}$$

Since $\frac{dB_v(t)}{dt}+\frac{dR_v(t)}{dt}=0$ holds for all $t\geq 0$ and all $v\in V$, $B_v(t)+R_v(t)=1$ for all $t$ and all $v\in V$. Therefore, we only need to consider the following master equation for $v\in V$:

$$\begin{aligned}\frac{dB_v(t)}{dt} &= f\left(\frac{1}{\deg(v,G_B)}\sum_{u\in N_{v,G_B}} B_u(t)\right)\Big[1-B_v(t)\Big] -\\ &\quad g\left(\frac{1}{\deg(v,G_R)}\sum_{u\in N_{v,G_R}} B_u(t)\right)B_v(t). \qquad (2)\end{aligned}$$

The main research task is to analyze system (2) for all $v\in V$.

**Remark.** When we investigate *specific* attacks and defenses, we need to obtain their concrete attack-power and defense-power functions. Similarly, when we investigate *specific* cyber systems, we need to obtain the concrete attack and defense network structures. These are important research problems that are orthogonal to the focus of the present paper because our *characterization study* deals with all possible attack-power and defense-power functions as well as all possible attack and defense network structures. In principle, these functions and structures do exist, although how to obtain them is an excellent problem for future investigation.

# 3. EQUILIBRIA AND THEIR STABILITY

Equilibrium is an important concept for quantifying cyber security. Suppose $\sigma$ is the equilibrium under certain active defense. We can quantify the effectiveness of active defense via the notion of *$\sigma$-effectiveness* because the dynamics converge to $\sigma$. Moreover, the *stability* of an equilibrium reflects the consequence/effect of perturbations, which can be caused (for example) by manipulations to the initial global state (e.g., the defender manually cleans up some compromised computers before launching active defense for more effectiveness — this may sound counterintuitive, but it actually shows the value of rigorous characterization study because the defender would not know this tactics otherwise).

We consider a class of equilibria of Eq. (2), namely homogeneous equilibria $[B_1^*,\cdots,B_n^*]$ with $B_1^*=\ldots=B_n^*=\sigma\in[0,1]$. This class contains the following:

- All-blue equilibrium, denoted by $B^*=1$; $B_v^*=1$ for all $v\in V$ (i.e., active defense is 1-*effective*).
- All-red equilibrium, denoted by $B^*=0$; $B_v^*=0$ for all $v\in V$ (i.e., active defense is 0-*effective*).
- $\sigma$-equilibrium, denoted by $B^*=\sigma\in(0,1)$; $B_v^*=\sigma$ for all $v\in V$ (i.e., active cyber defense is *$\sigma$-effective*).

The Jacobian matrix of (2) near an equilibrium is denoted by

$$\begin{aligned}M &= \Big[(1-\sigma)f'(\sigma)D_{A_B}^{-1}A_B-\sigma g'(\sigma)D_{A_R}^{-1}A_R\Big]-\\ &\quad \Big[f(\sigma)+g(\sigma)\Big]I_n. \qquad (3)\end{aligned}$$

## 3.1 Existence and Stability of Equilibria

We show that homogeneous equilibria exist under the following hypothesis (or condition):

> $\mathbf{H}_0$: there exists some $\sigma\in[0,1]$ such that $(1-\sigma)\cdot f(\sigma)=\sigma\cdot g(\sigma)$ holds.

PROPOSITION 1. *Under hypothesis $\mathbf{H_0}$, $B^*=\sigma\in[0,1]$ is an equilibrium of (2). Moreover, $B^*$ is stable if $\Re(\mu)<0$ for all $\mu\in\lambda(M)$, and unstable if $\Re(\mu)>0$ for some $\mu\in\lambda(M)$.*

PROOF. Under hypothesis $\mathbf{H_0}$, namely $(1-\sigma)\cdot f(\sigma)=\sigma\cdot g(\sigma)$, we see that $B_v^*=\sigma$ satisfies

$$\frac{dB_v(t)}{dt}=(1-\sigma)\cdot f(\sigma)-\sigma\cdot g(\sigma)=0,\quad \forall\, v\in V.$$

Thus $B^*=\sigma$ is an equilibrium.

To see the stability of equilibrium $B^*=\sigma\in[0,1]$, we consider a small perturbation to $B^*$, namely $\delta B=[B_1-B_1^*,\cdots,B_n-B_n^*]$. The linearization system of Eq. (2) near $B^*$ leads to

$$\begin{aligned}\frac{d\delta B}{dt} &= \Big\{\Big[(1-\sigma)f'(\sigma)D_{A_B}^{-1}A_B-\sigma g'(\sigma)D_{A_R}^{-1}A_R\Big]-\\ &\quad \Big[f(\sigma)+g(\sigma)\Big]I_n\Big\}\delta B, \qquad (4)\end{aligned}$$

where $I_n$ is the identity matrix of size $n$. Note that $M$ as defined in Eq. (3) is the coefficient matrix of linear system (4). The stability of equilibrium $B^*=\sigma$ is determined by the eigenvalues of matrix

$M$. For the general case $G_B = (V, E_B) \neq G_R = (V, E_R)$, it can be shown that

$$\lambda(M) = \lambda\Big((1-\sigma)f'(\sigma)D_{A_B}^{-1}A_B - \sigma g'(\sigma)D_{A_R}^{-1}A_R\Big) - \Big[f(\sigma)+g(\sigma)\Big]. \quad (5)$$

If $\Re(\mu) < 0$ for all $\mu \in \lambda(M_\sigma)$, $B^* = \sigma$ is locally stable; if $\Re(\mu) > 0$ for some $\mu \in \lambda(M)$, $B^* = \sigma$ is locally unstable. □

Proposition 1 can be simplified when $\sigma = 0$ and $\sigma = 1$.

COROLLARY 1. *If $g(1) = 0$, then $B^* = 1$ is an equilibrium. It is locally stable if $-g'(1) < f(1)$ and locally unstable if $-g'(1) > f(1)$.*

*If $f(0) = 0$, then $B^* = 0$ is an equilibrium. It is locally stable if $f'(0) < g(0)$ and locally unstable if $f'(0) > g(0)$.*

PROOF. To prove the first part, we observe that $g(1) = 0$ implies $\mathbf{H}_0$ holds for $\sigma = 1$, namely that $B^* = 1$ is an equilibrium of system (2). For $\sigma = 1$, it can be shown that Eq. (4) becomes

$$\frac{d\delta B}{dt} = = \Big[-g'(1)D_{A_R}^{-1}A_R - f(1)I_n\Big]\delta B.$$

Proposition 1 says that a sufficient condition under which equilibrium $B^* = 1$ is locally stable is

$$-g'(1)\Re(\mu) < f(1), \quad \forall\, \mu \in \lambda\Big(D_{A_R}^{-1}A_R\Big). \quad (6)$$

Since $g(1) = 0$ and $g(x) \geq 0$ for $x \in [0,1]$, $g(x)$ is *locally* non-increasing at $x = 1$ and thus $-g'(1) \geq 0$. Since the sum for every row in matrix $D_{A_R}^{-1}A_R$ equals 1, the Perron-Frobenius theorem [10] says that its largest eigenvalue is 1. From Eq. (6), we have

$$-g'(1)\Re(\mu) < -g'(1) < f(1), \quad \forall\, \mu \in \lambda\Big(D_{A_R}^{-1}A_R\Big).$$

That is, if $-g'(1) < f(1)$, then $B^* = 1$ is locally stable; if $-g'(1) > f(1)$, there exists at least one eigenvalue $\mu_0 \in \lambda\Big(D_{A_R}^{-1}A_R\Big)$, say $\mu_0 = 1$, such that $-g'(1)\Re(\mu_0) - f(1) > 0$, meaning that $B^* = 1$ is locally unstable.

To prove the second part, we observe that $f(0) = 0$ implies $\mathbf{H_0}$ with $\sigma = 0$, namely that $B^* = 0$ is an equilibrium of (2). For $\sigma = 0$, Eq. (4) becomes

$$\begin{aligned}\frac{d\delta B}{dt} &= \Big\{\Big[(1-0)\cdot f'(0)D_{A_B}^{-1}A_B - 0\cdot g'(0)D_{A_R}^{-1}A_R\Big] - \Big[f(0)+g(0)\Big]I_n\Big\}\delta B \\ &= \Big[f'(0)D_{A_B}^{-1}A_B - g(0)I_n\Big]\delta B.\end{aligned}$$

Proposition 1 says that the sufficient condition for equilibrium $B^* = 0$ to be locally stable is

$$f'(0)\Re(\mu) < g(0), \quad \forall\, \mu \in \lambda\Big(D_{A_B}^{-1}A_B\Big). \quad (7)$$

Since $f(0) = 0$ and $f(x) \geq 0$ for $x \in [0,1]$, $f(x)$ is *locally* non-decreasing at $x = 0$ and thus $f'(0) \geq 0$. Since the largest eigenvalue of $D_{A_B}^{-1}A_B$ is 1, from Eq. (7) we have

$$f'(0)\Re(\mu) < f'(0) < g(0), \quad \forall\, \mu \in \lambda\Big(D_{A_B}^{-1}A_B\Big).$$

That is, if $f'(0) < g(0)$, then $B^* = 0$ is locally stable; if $f'(0) > g(0)$, there exists at least one eigenvalue $\mu_0 \in \lambda\Big(D_{A_B}^{-1}A_B\Big)$, say $\mu_0 = 1$, such that $f'(0)\Re(\mu_0) - g(0) > 0$, meaning that $B^* = 0$ is locally unstable. □

In the special case $G_B = G_R$, namely $A_B = A_R$, we immediately obtain the following corollary of Proposition 1:

COROLLARY 2. *Suppose hypothesis $\mathbf{H_0}$ holds and $G_B = G_R = G$ (i.e., $A_B = A_R = A$). Let $\mu_1$ be the eigenvalue of $D_A^{-1}A$ that has the smallest real part. If the attack-power and defense-power functions satisfy one of the following two conditions:*
*(i). $(1-\sigma)f'(\sigma) - \sigma g'(\sigma) > 0$ and $\frac{f(\sigma)+g(\sigma)}{(1-\sigma)f'(\sigma)-\sigma g'(\sigma)} > 1$,*
*(ii). $(1-\sigma)f'(\sigma) - \sigma g'(\sigma) < 0$ and $\frac{f(\sigma)+g(\sigma)}{(1-\sigma)f'(\sigma)-\sigma g'(\sigma)} < \Re(\mu_1)$, then equilibrium $B^* = \sigma \in [0,1]$ is locally stable.*

*If the attack-power and defense-power functions satisfy one of the two following conditions:*
*(i). $(1-\sigma)f'(\sigma) - \sigma g'(\sigma) > 0$ and $\frac{f(\sigma)+g(\sigma)}{(1-\sigma)f'(\sigma)-\sigma g'(\sigma)} < 1$,*
*(ii). $(1-\sigma)f'(\sigma) - \sigma g'(\sigma) < 0$ and $\frac{f(\sigma)+g(\sigma)}{(1-\sigma)f'(\sigma)-\sigma g'(\sigma)} > \Re(\mu_1)$, then equilibrium $B^* = \sigma \in [0,1]$ is locally unstable.*

## 3.2 Examples

**Example 1: Stability effect of different defense-power functions vs. a fixed attack-power function.** Suppose $G_B = G_R$ is an Erdös-Rényi (ER) random graph instance $G = (V, E)$ with $|V| = 2{,}000$ and edge probability $p = 0.005$ (i.e., every pair of nodes is connected with probability 0.005, independent of each other). We consider attack-power function $g(x) = 1 - x$ against the following four scenarios of defense-power function $f(x)$:

- Scenario I: $f(x) = x^2$, meaning that $B^* = 0$ is stable and $B^* = 1$ is unstable.
- Scenario II: $f(x) = x^2 + x$, meaning that $B^* = 0$ is unstable and $B^* = 1$ is stable.
- Scenario III: $f(x) = x^2 + \frac{1}{2}x$, meaning that $B^* = 0$ and $B^* = 1$ are stable, but $B^* = \frac{1}{2}$ is unstable.
- Scenario IV: $f(x) = -2x^2 + 2x$, meaning that $B^* = \frac{1}{2}$ is stable, but $B^* = 0$ and $B^* = 1$ are unstable.

Figure 2 plots the phase portraits of $\langle B_v(t)\rangle = \frac{1}{|V|}\sum_{v\in V} B_v(t)$, the portion of secure nodes. We observe that the simulation results confirm the analytic results. Specifically, Figure 2(a) shows that $\langle B_v(t)\rangle$ converges to $B^* = 0$ when $B_v(0) < 1$ for all $v \in V$; Figure 2(b) shows that $\langle B_v(t)\rangle$ converges to $B^* = 1$ when $B_v(0) > 0$ for all $v \in V$; Figure 2(c) shows that $\langle B_v(t)\rangle$ converges to $B^* = 1$ when $B_v(0) > 0.5$ for all $v \in V$ and converges to $B^* = 0$ when $B_v(0) < 0.5$ for all $v \in V$; Figure 2(d) shows that $\langle B_v(t)\rangle$ converges to $B^* = 0.5$ when $0 < B_v(0) < 1$ for all $v \in V$.

| time $t$ | $f(x)$ | $g(x)$ | $B^*$ |
|---|---|---|---|
| $[0, 150]$ | $f(x) = x^2 + x$ | $g(x) = 1 - x$ | $B^* = 1$ |
| $[150, 300]$ | $f(x) = x^2$ | $g(x) = 1 - x$ | $B^* = 0$ |
| $[300, 400]$ | $f(x) = -2x^2 + 2x$ | $g(x) = 1 - x$ | $B^* = 0.5$ |
| $[400, 500]$ | $f(x) = x^2 + \frac{1}{2}x$ | $g(x) = 1 - x$ | $B^* = 1$ |

**Table 1: The dynamics go to the respective equilibrium $B^*$ under some combinations of defense-power function $f(x)$ and attack-power function $g(x)$.**

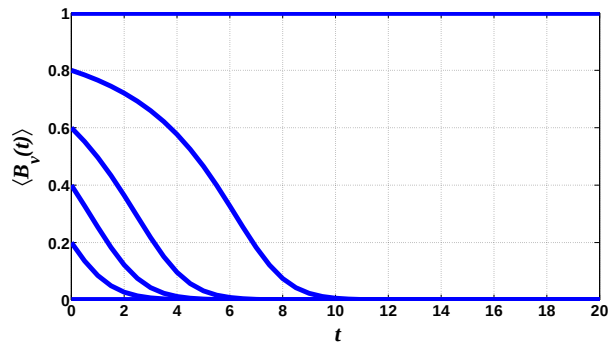

(a) Scenario I: $B^* = 0$ is stable, $B^* = 1$ is unstable.

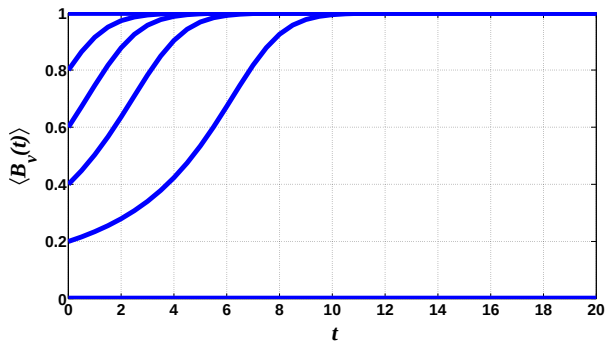

(b) Scenario II: $B^* = 0$ is unstable, $B^* = 1$ is stable.

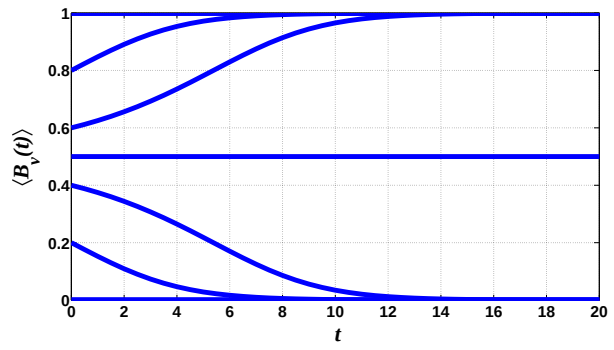

(c) Scenario III: $B^* = 0$ and $B^* = 1$ are stable, $B^* = \frac{1}{2}$ is unstable.

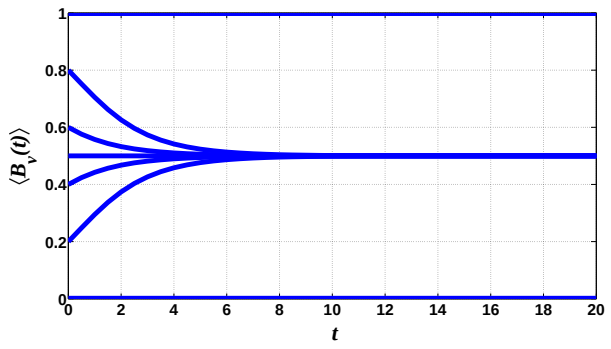

(d) Scenario IV: $B^* = 0$ and $B^* = 1$ are unstable, $B^* = \frac{1}{2}$ is stable.

**Figure 2: Phase portraits of the four scenarios confirming the stabilities of the equilibria, where $x$-axis represents time, and $y$-axis represents the portion of secure nodes $\langle B_v(t)\rangle$.**

Now we study the stability of the equilibria. For the $G_B = G_R$ mentioned above, we consider the above four scenarios as highlighted in Table 1. More specifically, for time $t \in [0, 150]$, the defense-power function is $f(x) = x^2 + x$ and the attack-power function is $g(x) = 1 - x$ (i.e, the above Scenario I); for time $t \in [150, 300]$, the defense-power function is $f(x) = x^2$ and the attack-power function is $g(x) = 1 - x$ (i.e., the above Scenario II); for time $t \in [300, 400]$, the defense-power function is $f(x) = -2x^2 + 2x$ and the attack-power function is $g(x) = 1 - x$ (i.e., the above Scenario IV); for time $t \in [400, 500]$, the defense-power function is $f(x) = x^2 + \frac{1}{2}x$ and the attack-power function is $g(x) = 1 - x$ (i.e., the above Scenario III).

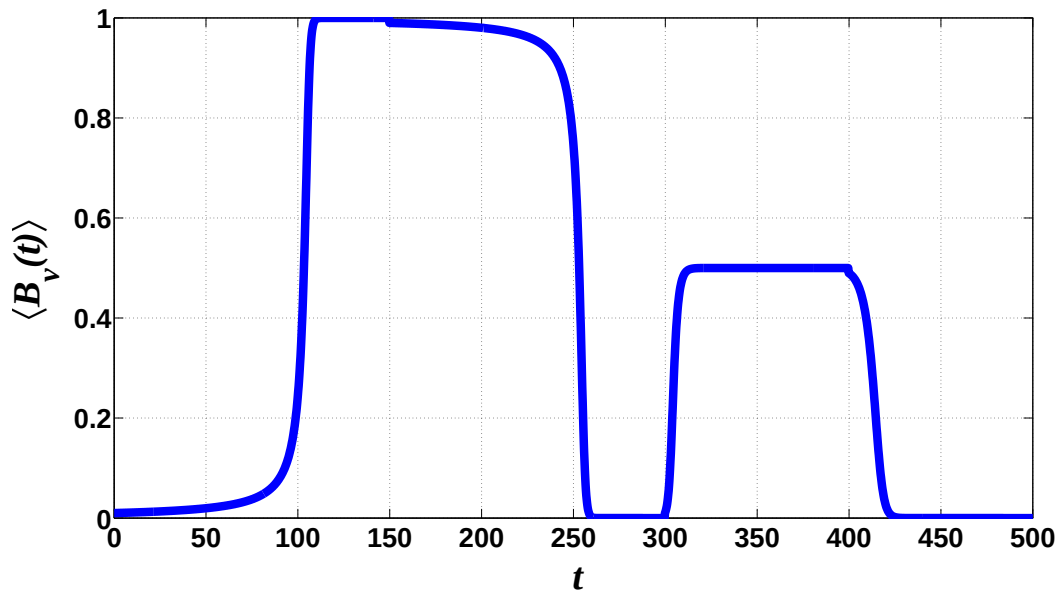

**Figure 3: Active cyber defense dynamics lack persistent equilibrium due to frequent perturbations.**

Figure 3 plots a very probable scenario that can happen to the portion of secure nodes, where three small perturbations are imposed at $t = 150, 300, 400$. This scenario is very probable because it can explain why the cyber security state may rarely enter some persistent equilibrium. Specifically, the initial value $B_v(0), v \in V$, is randomly chosen from interval $(0, 0.01]$ by the uniform distribution. At $t = 150$, we find that $\langle B_v(150)\rangle = 1$. We then impose a small perturbation on each $B_v(150)$, by replacing $B_v(150)$ with $B_v(150) - \varepsilon_v$ where $\varepsilon_v$ is an independent random variable of a uniform distribution in the interval $[0, 0.01]$ for all $v \in V$. Similarly, we replace $B_v(300)$ with $B_v(300) + \varepsilon_v$ and $B_v(400)$ with $B_v(400) - \varepsilon_v$ for all $v \in V$. Figure 3 illustrates that under small perturbations, the overall cyber security dynamics never enter any persistent equilibrium. This offers one possible explanation why real-life cyber security is perhaps never in any equilibrium.

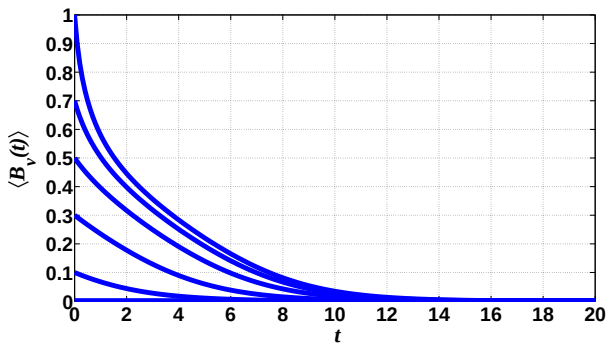

(a) $\nu = 0.5$

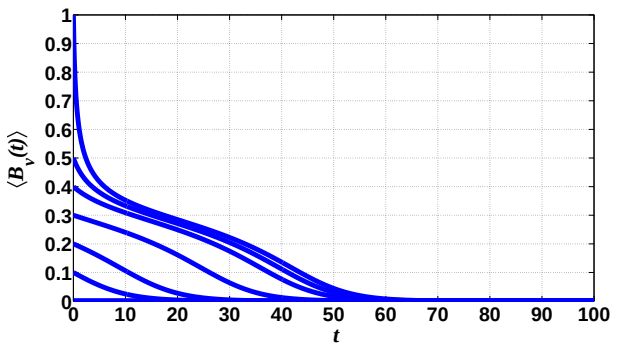

(b) $\nu = 0.8$

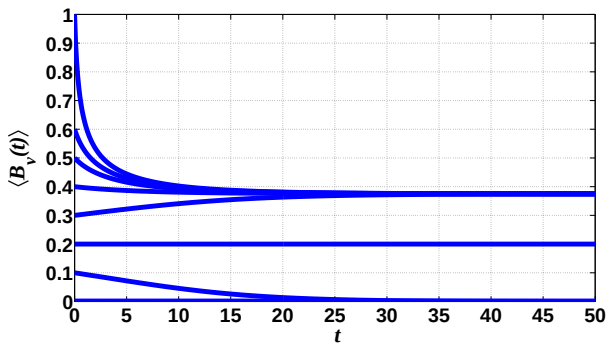

(c) $\nu = 0.85$

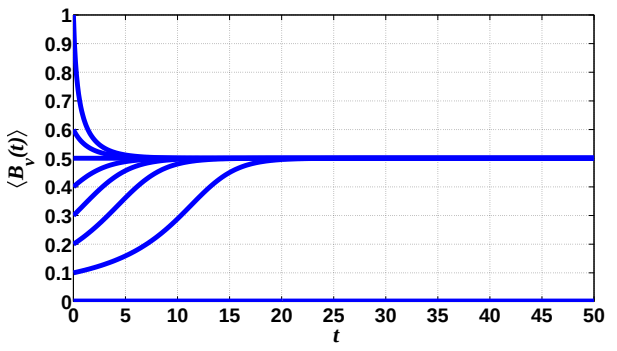

(d) $\nu = 1$

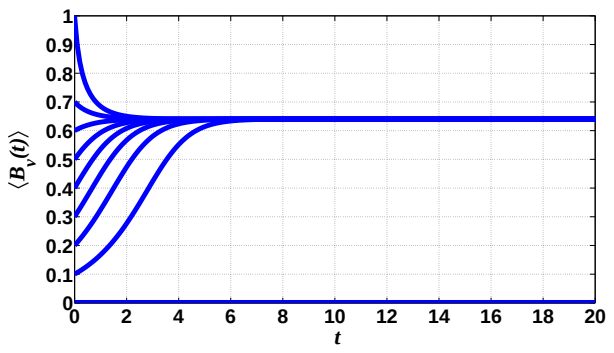

(e) $\nu = 1.5$

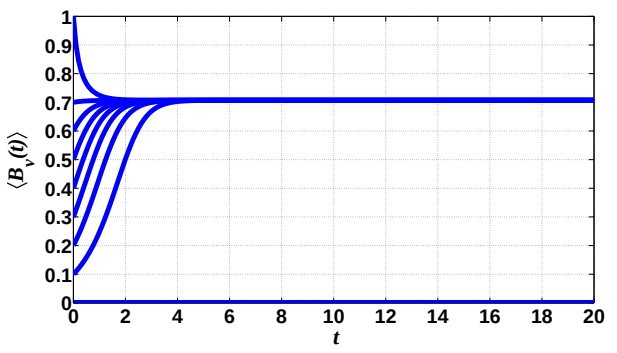

(f) $\nu = 2$

**Figure 4: Phase portraits of the portion of secure nodes $\langle B_v(t)\rangle$: $f(x,\nu) = \nu x - 2x^2$ and $g(x) = (1 - 2x)^2$.**

**Example 2: Stability effect of parameterized defense-power functions vs. a fixed attack-power function.** Suppose $G_B = G_R$ is an ER graph $G = (V, E)$ with $|V| = 2,000$, but with edge probability $p = 0.5$. We consider the following parameterized defense-power function $f(x, \nu)$ with parameter $\nu \in (0, +\infty)$ and fixed attack-power function $g(x)$:

$$f(x,\nu) = \nu x - 2x^2, \quad g(x) = (1 - 2x)^2.$$

Figure 4 plots the phase portraits of $\langle B_v(t)\rangle$ with $\nu = 0.5, 0.8, 0.85, 1, 1.5, 2$, respectively. The portraits can be classified into three classes. Figures 4(a)-4(b) show that there is one stable equilibrium $B^* = 0$. Figure 4(c) shows that there are three equilibria $B^* = 0, 0.38, 0.2$, where the first two are stable but the last one is unstable. Figures 4(d)-4(f) show that there exist two equilibria $B^* = 0, \sigma$ with $\sigma > 0$, where $B^* = 0$ is unstable and $B^* = \sigma$ is stable. We observe that active cyber defense dynamics exhibit different phenomena with respect to different parameters. Moreover, we observe a sort of phase transition in parameter $\nu$: when $\nu \le 0.8$, the global cyber security state converges to $B^* = 0$ almost regardless of the initial value; when $\nu \ge 1$, the global cyber security state converges to some $B^* = \sigma > 0$ almost regardless of the initial value; when $0.8 < \nu = 0.85 < 1$, the global cyber security state converges to some equilibrium dependent upon the initial value.

We summarize the discussion in this section into:

INSIGHT 1. *Active cyber defense dynamics may rarely enter into any equilibrium because of perturbations to the global security state as caused by the manual cleaning of some compromised computers (Figure 2), and/or because of perturbations to the attack/defense power function as caused by the introduction of a new attack/defense method (Figures 3-4 )*

## 4. TRANSITION BETWEEN MULTIPLE ATTRACTORS

We are now ready to precisely characterize the *transition* between the equilibria, which reflects the consequence/effect of the defender manipulating the initial global security state (e.g., manually cleaning up some compromised computers before launching active defense) and/or manipulating the attack/defense network structure (e.g., by changing the network access control policy to block/allow certain computers to communicate with certain other computers).

### 4.1 Transition Between the All-blue and All-red Equilibria

Under the conditions mentioned in Corollary 1, namely, $f(0) = g(1) = 0$, system (2) has two locally stable equilibria $B^* = 1$ and $B^* = 0$. Let $\mathbf{B} = \big[B_1, B_2, \cdots, B_n\big] \in [0,1]^n$ and $\mathbf{R} = \mathbf{1} - \mathbf{B} = \big[1 - B_1, 1 - B_2, \cdots, 1 - B_n\big] \in [0,1]^n$, where $n = |V|$. For $\tau_1^*, \tau_2^* \in (0,1)$, we define two sets $\Xi_{G_B,\tau_1^*}$ and $\Xi_{G_R,\tau_2^*}$ as follows:

$$\Xi_{G_B,\tau_1^*} = \left\{\mathbf{B} \in [0,1]^n \;\middle|\; \frac{1}{\deg(v,G_B)} \sum_{u\in N_{v,G_B}} B_u \geq \tau_1^*, \forall v \in V\right\}, \quad (8)$$

$$\Xi_{G_R,\tau_2^*} = \left\{\mathbf{R} \in [0,1]^n \;\middle|\; \frac{1}{\deg(v,G_R)} \sum_{u\in N_{v,G_R}} R_u \geq \tau_2^*, \forall v \in V\right\}. \quad (9)$$

The following Theorem 1, whose proof is deferred to the Appendix, gives the transition between the all-blue and all-red equilibria by manipulating the initial state $\mathbf{B}(0)$.

THEOREM 1. *Let $G_B = (V, E_B)$ and $G_R = (V, E_R)$ be two arbitrary graphs. Suppose that $f(\cdot)$ and $g(\cdot)$ are continuous with $f(0) = g(1) = 0$.*
**Case 1:** *Suppose the attack-power and defense-power functions satisfy, $\forall\, z \in [\tau_1^*, 1)$ and $\forall\, \mathbf{B} \in \Xi_{G_B,\tau_1^*}$ and some $\alpha > 0$,*

$$f(z) > \alpha \cdot z, \quad (10)$$

$$f\left(\frac{1}{\deg(v,G_B)} \sum_{u\in N_{v,G_B}} B_u\right) + g\left(\frac{1}{\deg(v,G_R)} \sum_{u\in N_{v,G_R}} B_u\right) \leq \alpha \quad (11)$$

*If initial value $\mathbf{B}(0) \in \Xi_{G_B,\tau_1^*}$, then $\lim_{t\to\infty} B_v(t) = 1\; \forall v \in V$.*
**Case 2:** *Suppose the attack-power and defense-power functions satisfy, $\forall\, z \in [\tau_2^*, 1)$ and $\forall\, \mathbf{R} \in \Xi_{G_R,\tau_2^*}$ and some $\beta > 0$, $g(1 - z) > \beta \cdot z$ and*

$$f\left(1 - \frac{1}{\deg(v,G_B)} \sum_{u\in N_{v,G_B}} R_u\right) + g\left(1 - \frac{1}{\deg(v,G_R)} \sum_{u\in N_{v,G_R}} R_u\right) \leq \beta \quad (12)$$

*If initial value $\mathbf{R}(0) \in \Xi_{G_R,\tau_2^*}$, then $\lim_{t\to\infty} R_v(t) = 1\; \forall v \in V$.*

The ***cyber security meaning*** of Theorem 1 is: Under a certain condition (**case 1**), the defender needs to manipulate the initial global security state $\mathbf{B}(0)$ to belong to $\Xi_{G_B,\tau_1^*}$ to make active defense *1-effective*; this says what the defender should strive to do. Under certain other circumstances (**case 2**), the defender should make sure that the initial global security state $\mathbf{B}(0)$ does not cause $\mathbf{R}(0) = \mathbf{1} - \mathbf{B}(0) \in \Xi_{G_R,\tau_2^*}$, because in this regime active defense is *0-effective*; this says what the defender should strive to avoid.

For the following two corollaries, we define

$$\Xi_{G_B,\tau^*} = \left\{\mathbf{B} \in [0,1]^n \;\middle|\; \frac{1}{\deg(v,G_B)} \sum_{u\in N_{v,G_B}} B_u > \tau^*, \forall\, v \in V\right\},$$

$$\Theta_{G_R,\tau^*} = \left\{\mathbf{B} \in [0,1]^n \;\middle|\; \frac{1}{\deg(v,G_R)} \sum_{u\in N_{v,G_R}} B_u < \tau^*, \forall\, v \in V\right\}.$$

On one hand, the following Corollary 3 says that when $\tau_1^* = \tau_2^* = \tau^*$, we obtain the same threshold for the transitions.

COROLLARY 3. *Suppose $f(\cdot)$ and $g(\cdot)$ are continuous with $f(0) = g(1) = 0$. There exist constants $\tau \in (0,1)$ and $\alpha > 0$ such that the following two conditions hold:*
*(i) The attack-power and the defense-power functions satisfy $f(z) > \alpha \cdot z$ for any $z \in (\tau^*, 1)$, and for any $\mathbf{B} \in \Xi_{G_B,\tau^*}$*

$$f\left(\frac{1}{\deg(v,G_B)} \sum_{u\in N_{v,G_B}} B_u\right) + g\left(\frac{1}{\deg(v,G_R)} \sum_{u\in N_{v,G_R}} B_u\right) \leq \alpha.$$

*(ii) The attack-power and the defense-power functions satisfy $g(z) > \alpha(1 - z)$ for any $z \in (0, \tau^*)$, and for any $\mathbf{B} \in \Theta_{G_R,\tau^*}$*

$$f\left(\frac{1}{\deg(v,G_B)} \sum_{u\in N_{v,G_B}} B_u\right) + g\left(\frac{1}{\deg(v,G_R)} \sum_{u\in N_{v,G_R}} B_u\right) \leq \alpha.$$

*If initial value $\mathbf{B}(0) \in \Xi_{G_B,\tau^*}$, then $\lim_{t\to\infty} B_v(t) = 1\; \forall\, v \in V$; if initial value $\mathbf{B}(0) \in \Theta_{G_R,\tau^*}$, then $\lim_{t\to\infty} B_v(t) = 0\; \forall\, v \in V$.*

On the other hand, the following Corollary 4 makes a connection to [36], by accommodating Theorems 1, 5, 8 and 9 in [36] as a special case with $G_B = G_R$ and $\alpha = 1$.

COROLLARY 4. *Suppose $G_B = G_R = G = (V, E)$ and $f(\cdot)$, and $g(\cdot)$ are continuous with $f(0) = g(1) = 1$. There exist $\tau^* \in (0,1)$ and $\alpha > 0$ such that the attack-power and defense-power functions satisfy*

$$f(z) + g(z) \leq \alpha \;\; \forall z \in [0,1]$$

*and the defense-power function satisfy*

$$f(z) > \alpha \cdot z\; \forall z \in (\tau^*, 1) \text{ and } f(z) < \alpha \cdot z\; \forall z \in (0, \tau^*).$$

*If initial value $\mathbf{B}(0) \in \Xi_{G,\tau^*}$, then $\lim_{t\to\infty} B_v(t) = 1$ for all $v \in V$; if initial value $\mathbf{B}(0) \in \Theta_{G,\tau^*}$, then $\lim_{t\to\infty} B_v(t) = 0$ for all $v \in V$.*

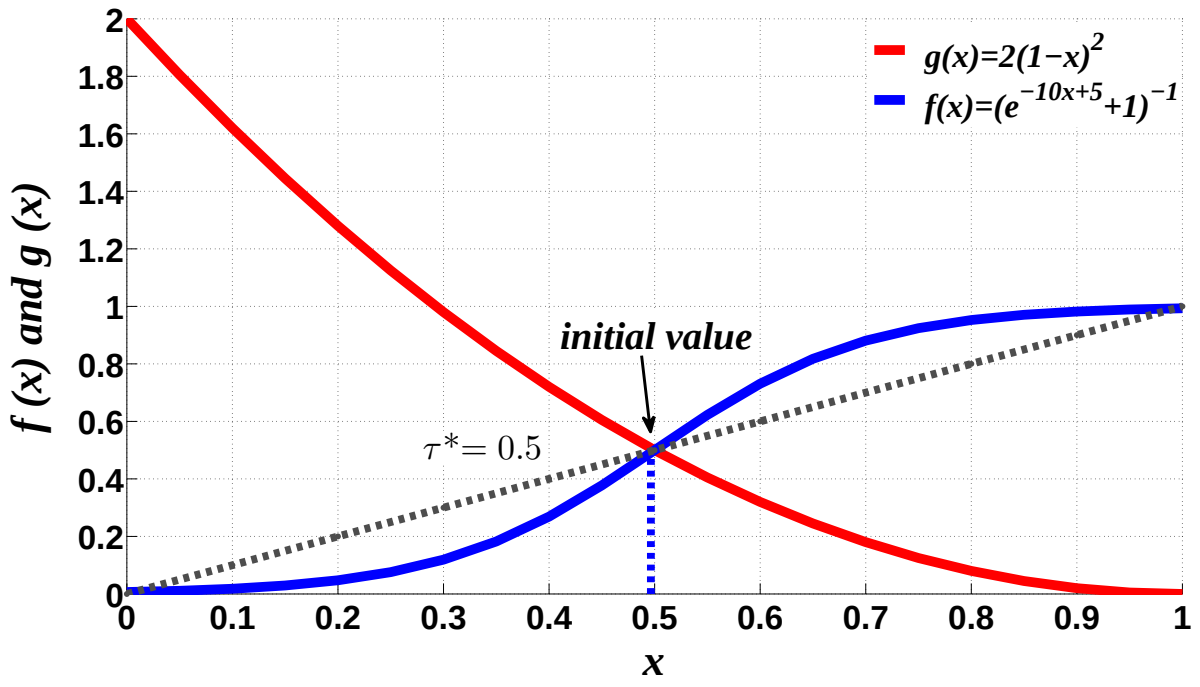

(a) $f(x)$, $g(x)$ and threshold $\tau^* = 0.5$ satisfy the condition of transition between $B^* = 0$ and $B^* = 1$

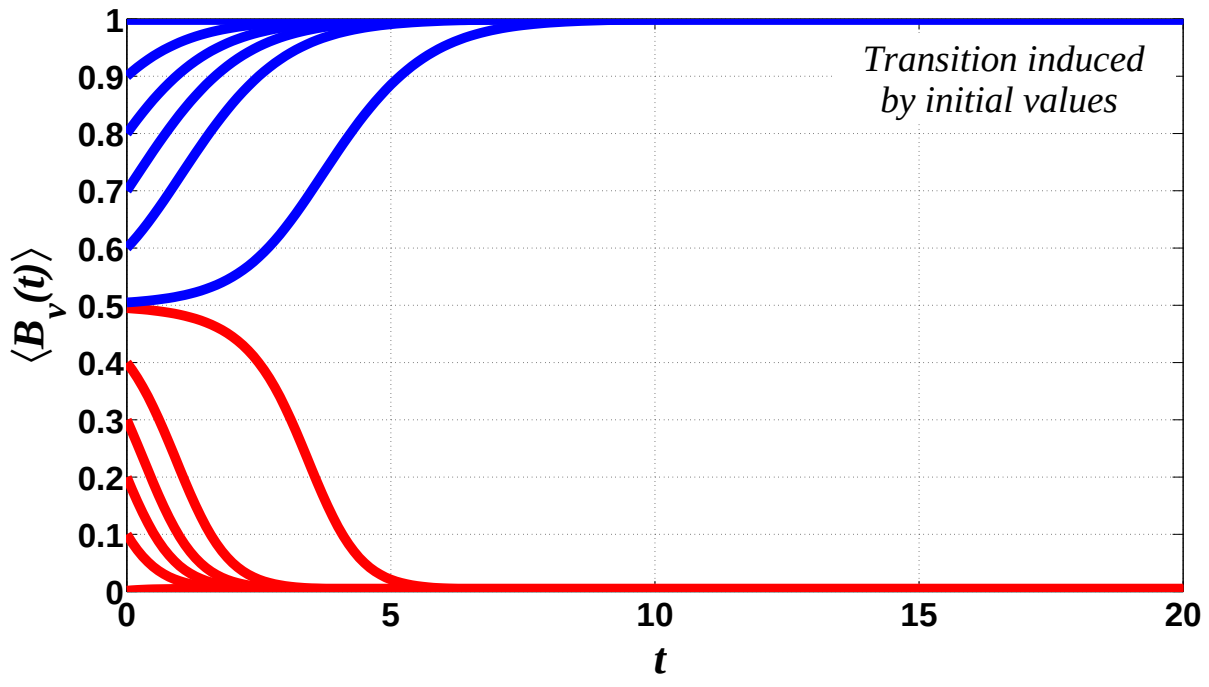

(b) Transition induced by varying initial value $\langle B_v(0)\rangle$

**Figure 5: Transition between equilibria $B^* = 0$ and $B^* = 1$ as induced by varying the initial value.**

## 4.2 Example

We consider the transition between equilibria $B^* = 0$ and $B^* = 1$ as caused by varying the initial value $\mathbf{B}(0)$. We use two concrete defense-power and attack-power functions:

$$f(x) = \frac{1}{e^{-10x+5}+1}, \quad g(x) = 2(1-x)^2,$$

which are plotted in Figure 5(a). The graphs $G_B$ and $G_R$ are two ER graph instances with $|V| = 2,000$ and $p = 0.5$. We consider the transition induced by varying the initial value $\langle B_v(0)\rangle$ between 0 and 1. Figure 5(b) shows that when $\langle B_v(0)\rangle > 0.5$, the dynamics converge to $B^* = 1$; when $\langle B_v(0)\rangle < 0.5$, the dynamics converge to $B^* = 0$.

The exploration in this section can be summarized as:

INSIGHT 2. *A small change in the initial global security state, in the model parameters, in the attack network structure, or in the defense network structure can lead to substantial change in active cyber defense dynamics. A rigorous characterization, such as Theorem 1, can offer precise guidance on "what the defender should strive to do" and "what the defender should strive to avoid" (e.g., how to manipulate the dynamics to benefit the defender rather than the attacker).*

# 5. HOPF BIFURCATION

We consider Hopf bifurcation near equilibrium $B^* = \sigma \in (0,1)$ under condition $(1-\sigma)\cdot f(\sigma) = \sigma \cdot g(\sigma)$. Recall that the stability of $B^* = \sigma \in (0,1)$ depends on $\lambda_1(M)$, where $M$, as defined in Eq. (3), is the Jacobian matrix of system (2). In the rest of the paper, we may simplify the notation $\lambda_1(M)$ as $\lambda_1$ unless there is potential ambiguity.

Consider differentiable defense-power and attack-power functions $f(x,\nu)$ and $g(x,\nu)$ with parameter $\nu$. Suppose $\frac{\partial f}{\partial \nu}$, $\frac{\partial g}{\partial \nu}$ and $\frac{\partial M}{\partial \nu}$ all depend on $\nu$. Consider the following critical condition for Hopf bifurcation:

$$\Re(\lambda_1) = 0 \ \text{ and } \ \Im(\lambda_1) \neq 0. \tag{13}$$

It is known that if (13) holds for some $\nu = \nu^*$, $\lambda_1(\nu)$ is differentiable in $\nu$, and $\frac{d\lambda_1}{d\nu} \neq 0$ at $\nu = \nu^*$, then system (2) exhibits Hopf bifurcation [24]. Therefore, we need to find the critical value $\nu^*$. For this purpose, we adopt the approach described in [20] to investigate how $\lambda_1$ depends on the permutation to $M$, namely to conduct a perturbation spectral analysis to compute the perturbation to $\lambda_1$, denoted by $\delta\lambda_1$, as caused by perturbation to $M$, denoted by $\delta M$.

## 5.1 How to Estimate $\delta\lambda_1$

Let $\mathbf{x}_1$ be the eigenvector of $M$ associated to eigenvalue $\lambda_1$, namely, $M\mathbf{x}_1 = \lambda_1\mathbf{x}_1$. For perturbation $\delta M$ to $M$, $M + \delta M$ can be described as $M(\nu) + M'(\nu)\delta\nu$. The perturbation to $M$ causes perturbation $\delta\lambda_1$ to $\lambda_1$ and perturbation $\delta\mathbf{x}_1$ to $\mathbf{x}_1$. That is,

$$\big(M + \delta M\big)\big(\mathbf{x}_1 + \delta\mathbf{x}_1\big) = \big(\lambda_1 + \delta\lambda_1\big)\big(\mathbf{x}_1 + \delta\mathbf{x}_1\big).$$

By ignoring the second-order term, we obtain

$$M\delta\mathbf{x}_1 + \delta M\mathbf{x}_1 = \lambda_1\delta\mathbf{x}_1 + \delta\lambda_1\mathbf{x}_1. \tag{14}$$

By multiplying both sides of Eq. (14) with the left eigenvector $\mathbf{y}_1$ corresponding to $\lambda_1$, we obtain

$$\mathbf{y}_1^\top M\delta\mathbf{x}_1 + y_1^\top\delta M\mathbf{x}_1 = \mathbf{y}_1^\top\lambda_1\delta\mathbf{x}_1 + y_1^\top\delta\lambda_1\mathbf{x}_1,$$

$$\mathbf{y}_1^\top\lambda_1\delta\mathbf{x}_1 + \mathbf{y}_1^\top\delta M\mathbf{x}_1 = \mathbf{y}_1^\top\lambda_1\delta\mathbf{x}_1 + \mathbf{y}_1^\top\delta\lambda_1\mathbf{x}_1,$$

$$\mathbf{y}_1^\top\delta M\mathbf{x}_1 = \mathbf{y}_1^\top\delta\lambda_1\mathbf{x}_1.$$

As a result, we can estimate $\delta\lambda_1$ as

$$\delta\lambda_1 = \frac{\mathbf{y}_1^\top\delta M\mathbf{x}_1}{\mathbf{y}_1^\top\mathbf{x}_1}, \tag{15}$$

where $\delta M$ can be estimated depending on whether the perturbation is to the attack and/or defense power (**Case A** below) or to the attack/defense network structure (**Case B** below).

**Case A: $\delta M$ is caused by perturbation to attack- and/or defense power.** Suppose the perturbation is imposed on parameter $\nu$ in the attack-power and defense-power functions $f(x,\nu)$ and $g(x,\nu)$, where $\frac{\partial f}{\partial \nu}$ and $\frac{\partial g}{\partial \nu}$ depend on $\nu$ as mentioned above. The cyber security meanings of such perturbations is (for example) that new attack and/or defense techniques are introduced. Note that

$$\begin{aligned}&\delta M(\nu)\\ =\ &\left\{\left[(1-\sigma)\frac{\partial f'(\sigma,\nu)}{\partial\nu}D_{A_B}^{-1}A_B - \sigma\frac{\partial g'(\sigma,\nu)}{\partial\nu}D_{A_R}^{-1}A_R\right] - \right.\\ &\left.\left[\frac{\partial f(\sigma,\nu)}{\partial\nu} + \frac{\partial g(\sigma,\nu)}{\partial\nu}\right]I_n\right\}\delta\nu.\end{aligned}$$

In the special case $G_B = G_R = G$ (i.e., the adjacency matrix $A_B = A_R = A$), we have

$$M = \big[(1-\sigma)f'(\sigma) - \sigma g'(\sigma)\big]D_A^{-1}A - \big[f(\sigma) + g(\sigma)\big]I_n,$$

the eigenvalues of $M$ are $\big[(1-\sigma)f'(\sigma) - \sigma g'(\sigma)\big]\mu - \big[f(\sigma) + g(\sigma)\big]I_n$ for all $\mu \in \lambda(D_A^{-1}A)$, and the perturbation can be rewritten as

$$\delta M(\nu) = \left\{\left[(1-\sigma)\frac{\partial f'(\sigma,\nu)}{\partial \nu} - \sigma\frac{\partial g'(\sigma,\nu)}{\partial \nu}\right]D_A^{-1}A - \left[\frac{\partial f(\sigma,\nu)}{\partial \nu} + \frac{\partial g(\sigma,\nu)}{\partial \nu}\right]I_n\right\}\delta\nu.$$

Hence, (15) becomes

$$\delta\lambda_1 = \mathbf{y}_1^\top\left\{\left[(1-\sigma)\frac{\partial f'(\sigma,\nu)}{\partial \nu} - \sigma\frac{\partial g'(\sigma,\nu)}{\partial \nu}\right]D_A^{-1}A - \left[\frac{\partial f(\sigma,\nu)}{\partial \nu} + \frac{\partial g(\sigma,\nu)}{\partial \nu}\right]I_n\right\}\delta\nu \cdot \mathbf{x}_1 \Big/ \mathbf{y}_1^\top\mathbf{x}_1. \quad (16)$$

**Case B: $\delta M$ is caused by perturbation to attack and/or defense network structure.** Suppose the perturbation is imposed on $G_B = (V, E_B)$ and/or $G_R = (V, E_R)$ by adding/deleting edges. The cyber security meaning of such perturbations is that the network is disrupted (e.g., edges are deleted by the attacker, or security policies have changed) and then edges are added by the defender. We assume that the number of added/deleted edges is small (compared with $|E_B|$ and $|E_R|$, respectively) so that we can approximately treat $\delta M$ as a small perturbation. Let $C_B = D_{A_B}^{-1}A_B$ and $C_R = D_{A_R}^{-1}A_R$. Perturbations to $A_B$ and $A_R$ lead to $A_B + \delta A_B$ and $A_R + \delta A_R$, respectively. Correspondingly, we obtain the perturbations to $C_B$ and $C_R$:

$$\begin{aligned}\delta C_B &= D_{A_B+\delta A_B}^{-1}(A_B + \delta A_B) - D_{A_B}^{-1}A_B,\\ \delta C_R &= D_{A_R+\delta A_R}^{-1}(A_R + \delta A_R) - D_{A_R}^{-1}A_R.\end{aligned}$$

Then, the perturbation to Jacobian matrix $M$ is

$$\delta M = (1-\sigma)f'(\sigma)\delta C_B - \sigma g'(\sigma)\delta C_R.$$

From (15), we have

$$\delta\lambda_1 = \frac{\mathbf{y}_1^\top\Big[(1-\sigma)f'(\sigma)\delta C_B - \sigma g'(\sigma)\delta C_R\Big]\mathbf{x}_1}{\mathbf{y}_1^\top\mathbf{x}_1}.$$

Note that in the special case $G_B = G_R = G$ (i.e., $A_B = A_R = A$) with perturbations $\delta C_B = \delta C_R$, we have

$$\begin{aligned}\delta M &= \Big[(1-\sigma)f'(\sigma) - \sigma g'(\sigma)\Big]\delta C,\\ \delta\lambda_1 &= \frac{\mathbf{y}_1^\top\Big[(1-\sigma)f'(\sigma) - \sigma g'(\sigma)\Big]\delta C\mathbf{x}_1}{\mathbf{y}_1^\top\mathbf{x}_1}.\end{aligned}$$

## 5.2 Example: Hopf Bifurcation Induced by Perturbation to Parameter

In order to show that Hopf bifurcation can happen, we consider an ER graph $G_B = G_R = G = (V, E)$ with $|V| = 2,000$ and edge probability $p = 0.005$. Let $\mu_1$ denote the eigenvalue of $D_A^{-1}A$ with the smallest real part, where $A$ is the adjacency matrix of $G$. For the ER graph, we have $\Re(\mu_1) = -0.3448$. We consider the following defense-power and attack-power functions:

$$f(x) = -4x^2 + 4x, \;\; g(x,\nu) = \Big(\nu x - \frac{\nu}{2}\Big)^2,$$

where $f(x)$ does not depend on $\nu$. Recall that under condition $(1-\sigma)f(\sigma) = \sigma g(\sigma)$, there exists equilibrium $B^* = \sigma \in (0,1)$.

When $\nu = 3$, we have homogeneous equilibrium $B^* = 0.7$, which is locally stable according to the second condition in the first part of Corollary 2:

$$\begin{aligned}(1-\sigma)f'(\sigma) - \sigma g'(\sigma,3) &= -3 < 0,\\ \frac{f(\sigma) + g(\sigma,3)}{(1-\sigma)f'(\sigma) - \sigma g'(\sigma,3)} &= -0.4 < \Re(\mu_1) = -0.3448.\end{aligned}$$

When $\nu = 4$, we have homogeneous equilibrium $B^* = 0.6667$, which is locally unstable according to the second condition in the second part of Corollary 2:

$$\begin{aligned}(1-\sigma)f'(\sigma) - \sigma g'(\sigma,4) &= -4 < 0\\ \frac{f(\sigma) + g(\sigma,4)}{(1-\sigma)f'(\sigma) - \sigma g'(\sigma,4)} &= -0.3333 > \Re(\mu_1) = -0.3448.\end{aligned}$$

Therefore, there is a *critical value* between $\nu = 3$ and $\nu = 4$, at which $\Re(\lambda_1(M)) = 0$. By conducting 100 independent simulation runs of $\nu \in [3,4)$ with step-length 0.01, we find the critical value $\nu = 3.8$ and the corresponding equilibrium $B^* = 0.6724$, where

$$\begin{aligned}(1-\sigma)f'(\sigma) - \sigma g'(\sigma,3.8) &= 7 - 3.81 < 0,\\ \frac{f(\sigma) + g(\sigma,3.8)}{(1-\sigma)f'(\sigma) - \sigma g'(\sigma,3.8)} &= -0.3448 = \Re(\mu_1).\end{aligned}$$

Figure 6(a) plots the periodic trajectory of $\langle B_v(t)\rangle$ when $\nu = 4 > 3.8$, which surrounds equilibrium $B^* = 0.6724$. Figure 6(b) plots the periodic trajectory of $\langle B_v(t)\rangle$ when $\nu = 5.05 > 3.8$. Figure 6(c) plots the bifurcation diagram with respect to $\nu \in (3,6)$. Figure 6(d) plots the bifurcation diagram with respect to $\nu \in (4.75, 5.5)$. We observe that when $\nu \in (5, 5.5)$, there are not only two-periodic trajectories, but also $k$-periodic trajectories ($k > 2$). In summary, the periodic trajectories exhibit the *period-doubling cascade* phenomenon.

## 5.3 Example: Hopf Bifurcation Induced by Perturbation to Attack/Defense Network Structures

For the purpose of demonstrating the bifurcation phenomenon caused by perturbation to network structures, we use two randomly generated ER graph examples $G_B = (V, E_B)$ and $G_R = (V, E_R)$, both with $|V| = 2,000$ and $p = 0.005$. The average degree is 10.0565 for $G_B$ and 11.1865 for $G_R$. We use the following defense-power and attack-power functions:

$$f(x) = -4x^2 + 4x, \;\; g(x,\nu) = \Big(\nu x - \frac{\nu}{2}\Big)^2 \;\; \text{with} \;\; \nu = 6$$

We perform 100 iterations of operations to $G_R$ as follows: during each of the first 50 iterations, we delete 226 edges (or 1% of the edges in the original $E_R$) chosen independently and uniformly at random; during each of the following 50 iterations, we add 226 edges chosen independently and uniformly random among all the unconnected edges. That is, we delete and then add 50% edges of the original $|E_R|$.

Figure 7 demonstrates that the *period-doubling cascade* phenomenon appears and finally leads to chaos after deleting more than 36% edges and before adding 14% edges. We observe that eventually the diagram becomes stable after adding the same number of edges as those deleted. (Note that Figure 7 is not symmetric because the added edges are random and in general are different from the edges that are deleted.)

The following insight summarizes the exploration of this section.

INSIGHT 3. *Active cyber defense dynamics can exhibit Hopf bifurcation, when the attack/defense power varies in certain parameter regimes and/or when the attack/defense network structure varies*

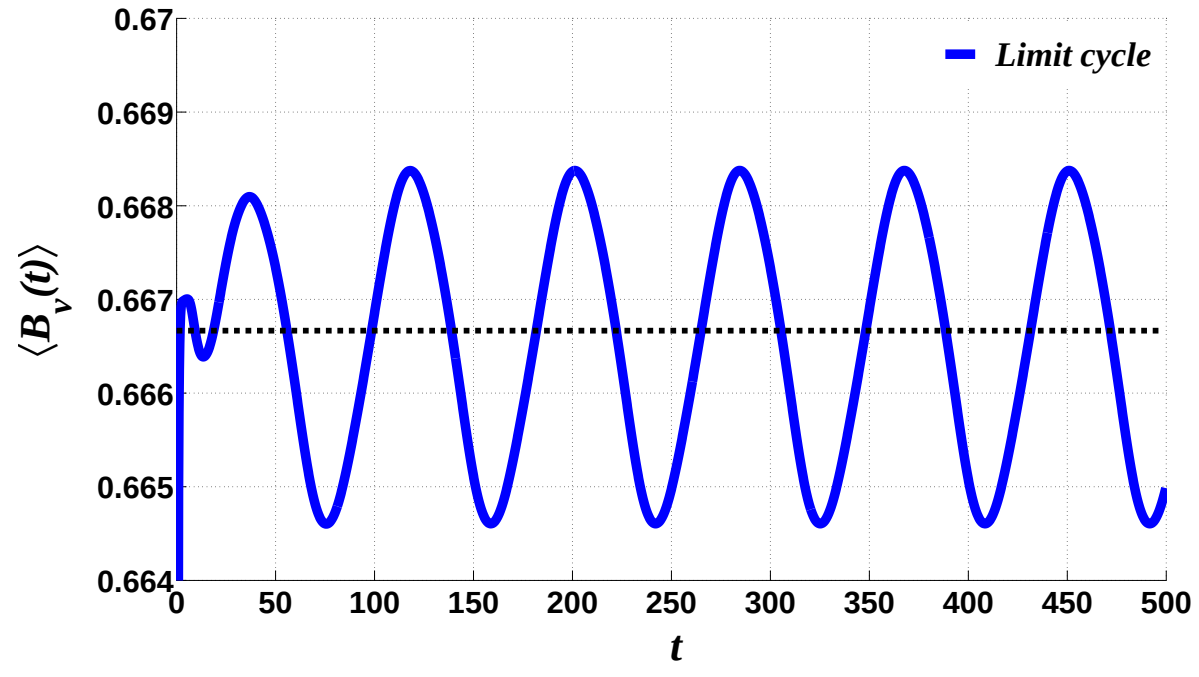

(a) Periodic trajectory of $\langle B_v(t)\rangle$ with $\nu = 4$

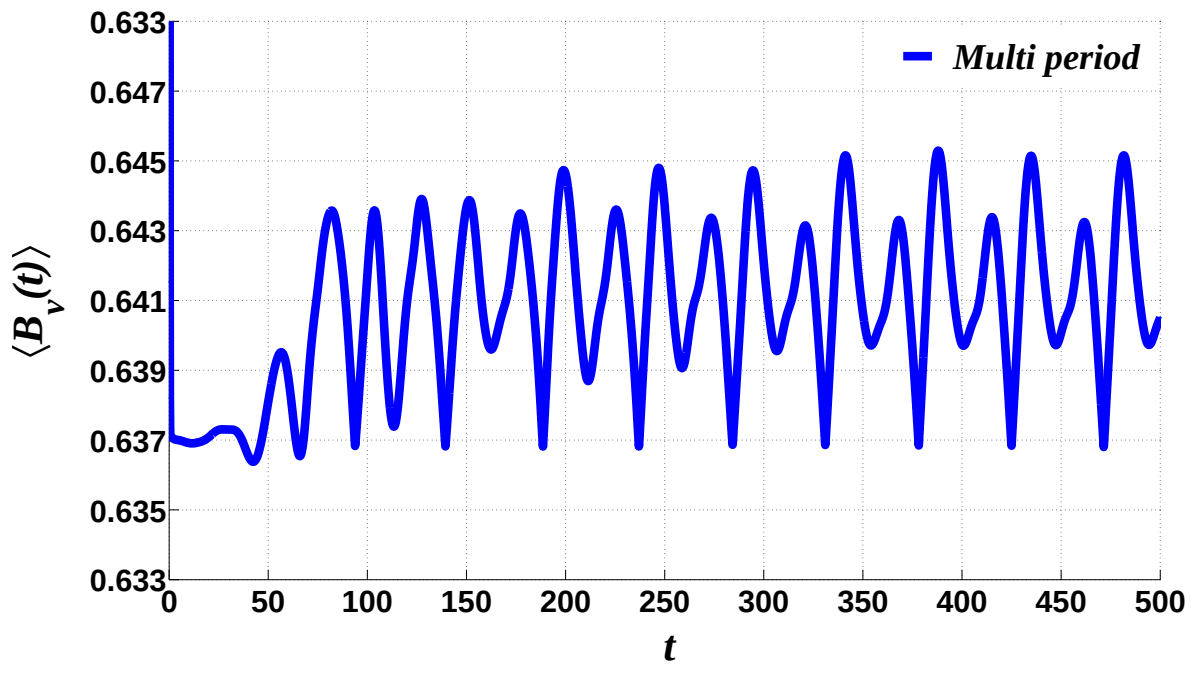

(b) Periodic trajectory of $\langle B_v(t)\rangle$ w/ $\nu = 5.05$

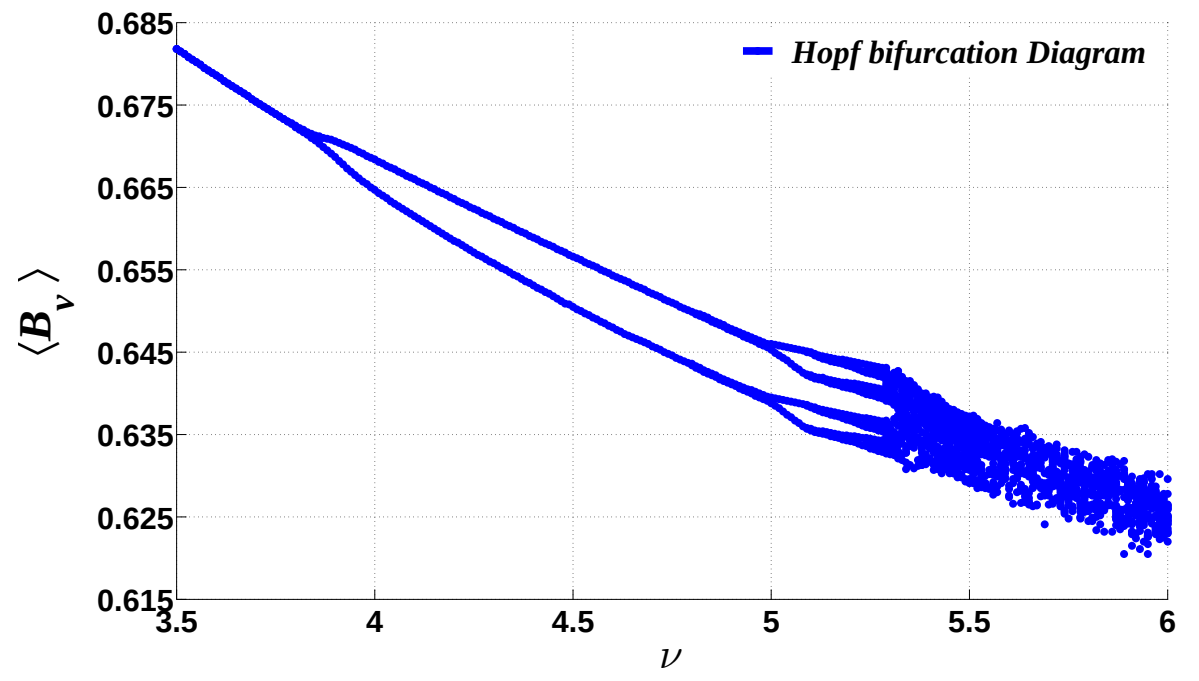

(c) Bifurcation diagram w/ $\nu \in (3, 6)$

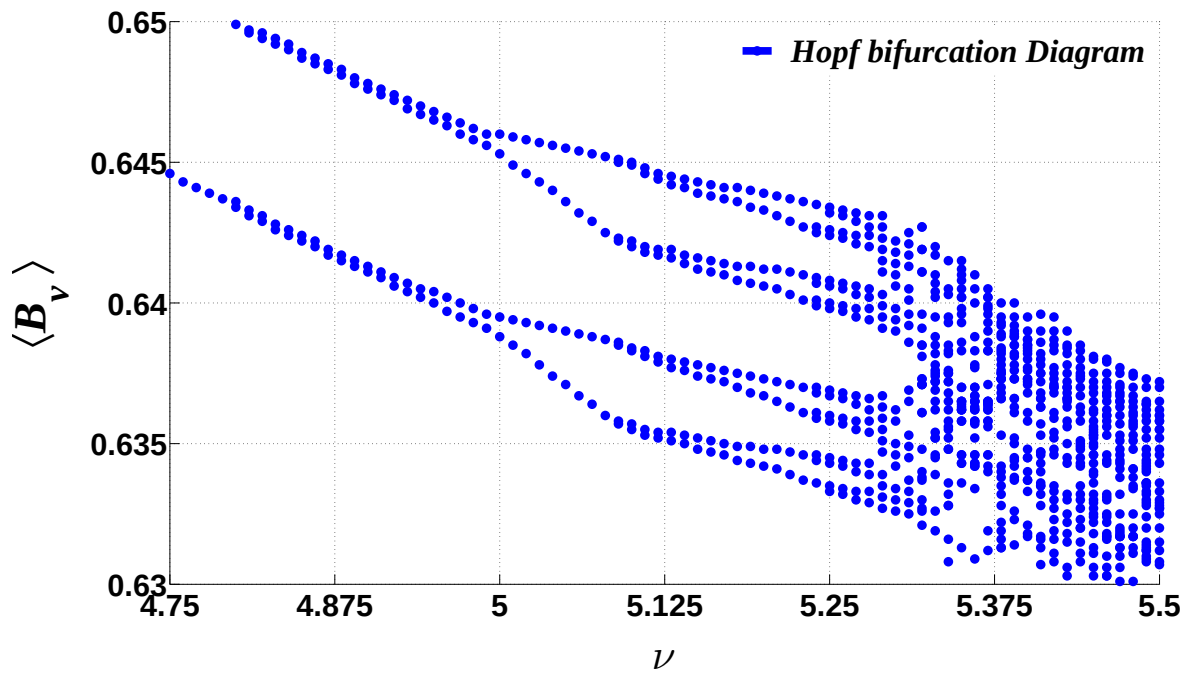

(d) Bifurcation diagram w/ $\nu \in (4.75, 5.5)$

**Figure 6: Limit cycle and Hopf bifurcation diagram, where $\langle B_v\rangle$ are the extremum points of $\langle B_v(t)\rangle$ in time period $t \in (1000, 2000)$.**

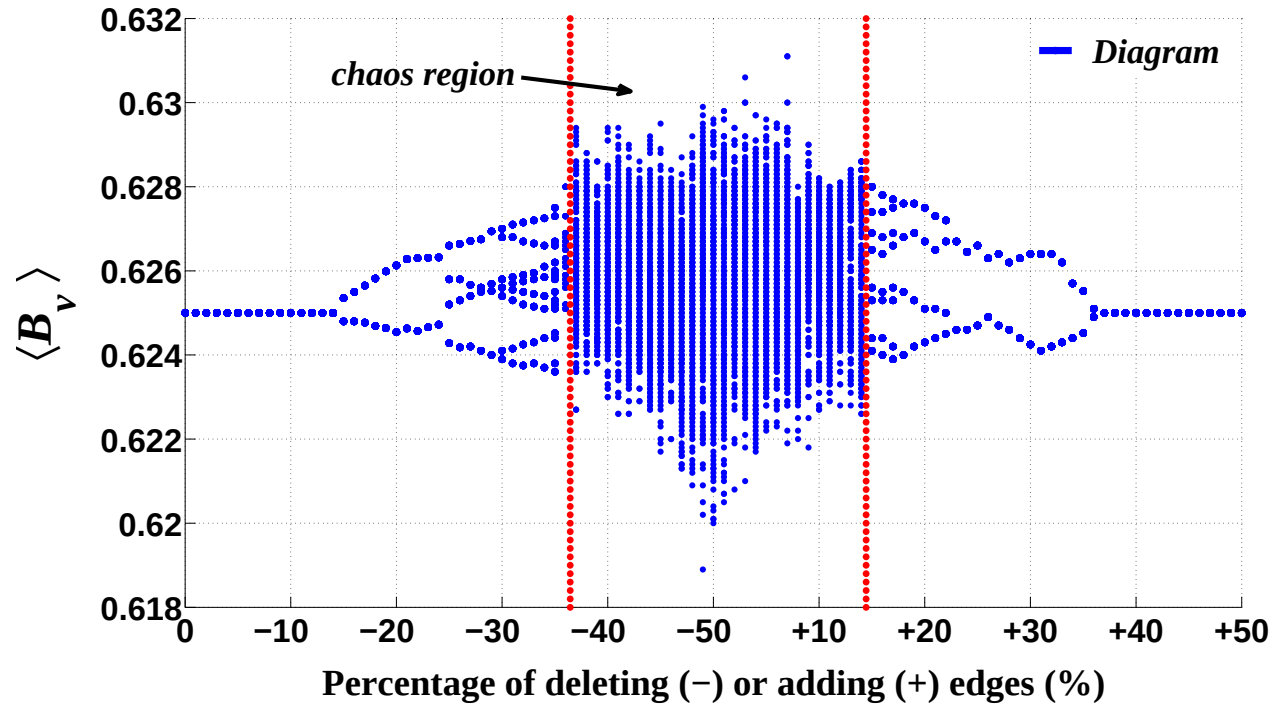

**Figure 7: Hops bifurcation induced by perturbation to the network structure.**

*in certain patterns. These situations are "unmanageable" because it would be infeasible, if not impossible, to estimate the global security state in real-time. Therefore, the defender must strive to avoid such unmanageable situations by manipulating the dynamics carefully (e.g., by disrupting the bifurcation condition or containing the attack-power of the adversary).*

# 6. CHAOS

Figure 6(c) shows that the number of periodic points increase with parameter $\nu$, which hints that system (2) can exhibit the chaos phenomenon. To see this, we consider the case $G_B = G_R$. In this case, system (2) becomes

$$\begin{aligned}\frac{dB_v(t)}{dt} &= f\left(\frac{1}{\deg(v,G)}\sum_{u\in N_{v,G}} B_u(t)\right)\Big[1 - B_v(t)\Big] - \\ & g\left(\frac{1}{\deg(v,G)}\sum_{u\in N_{v,G}} B_u(t)\right) B_v(t).\end{aligned}$$

Let $F\big(B_v(0), t\big)$ denote the right-hand part. Consider $B_v(0)$ and $B_v(0) + \varepsilon_v(0)$ for all $v \in V$, where $\varepsilon_v(0) \in \mathbb{R}^n$ is a small perturbation to the initial point $B_v(0)$. Then, we have $\forall v \in V$,

$$\begin{aligned}\varepsilon_v(t) &= F\big(B_v(0) + \varepsilon_v(0), t\big) - F\big(B_v(0), t\big) \\ &= DF\big(B_v(0), t\big) \cdot \varepsilon_v(0),\end{aligned}$$

where $DF\big(B_v(0), t\big)$ is the Jacobian matrix of the map $F$ at time $t$. By the QR decomposition of matrix $\varepsilon(t) = [\varepsilon_1(t), \varepsilon_2(t), \cdots, \varepsilon_n(t)]$ where $n = |V|$, we obtain matrix

$$\varepsilon(t) = q(t) \cdot r(t),$$

where $q(t)$ is an orthogonal matrix and $r(t)$ is an upper triangular matrix. Note that $\varepsilon(t) = q(t)$ and the diagonal element $\lambda_{ii}(t)$ of $r_t$ at time $t$ is the exponential magnification, where $i \in \{1, 2, \cdots, n\}$. Thus, the average rate of divergence or convergence of the two trajectories $\big\{F\big(B_v(0), t\big)\big|t \geq 0\big\}$ and $\big\{F\big(B_v(0)+\varepsilon_v(0), t\big)\big|t \geq 0\big\}$ for all $v \in V$ is defined by

$$L_i = \lim_{t\to\infty} \frac{1}{t} \ln \lambda_{ii}(t),$$

where $L_i$ for $i = 1, 2, \cdots, n$ are the Lyapunov characteristic exponents. It is known [24] that under some mild conditions, the above limit exists and is finite for almost all initial values $\mathbf{B}(0) = [B_1(0), B_2(0), \cdots, B_n(0)]$ and for almost all matrices $\varepsilon(0)$. Note that $\mathsf{MLE} = \max_{1\leq i\leq n} L_i$ indicates whether the dynamical system is chaotic or not. More specifically, when $\mathsf{MLE} > 0$, a small perturbation to the initial value will lead to an exponential separation and therefore leads to the chaos phenomenon.

**Example.** Consider an ER graph instance $G_B = G_R$ with $|V| = 2,000$ and $p = 0.005$, and the following defense-power and attack-power functions:

$$f(x) = -4x^2 + 4x, \;\; g_\nu(x) = \left(\nu x - \frac{\nu}{2}\right)^2.$$

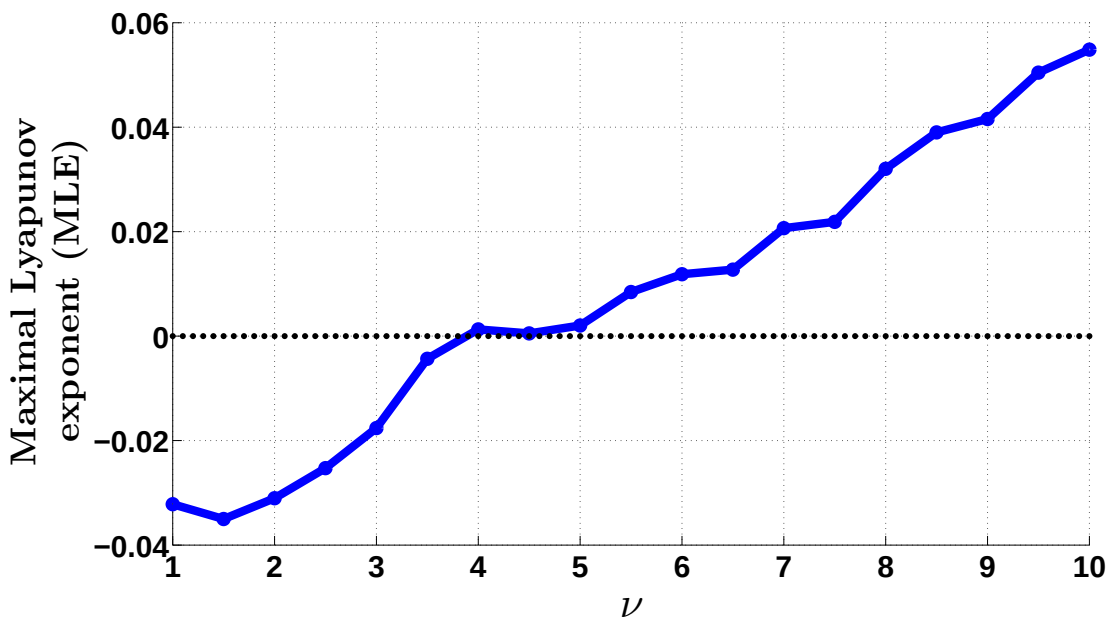

(a) MLE with $\nu$: MLE $> 0$ indicates chaos.

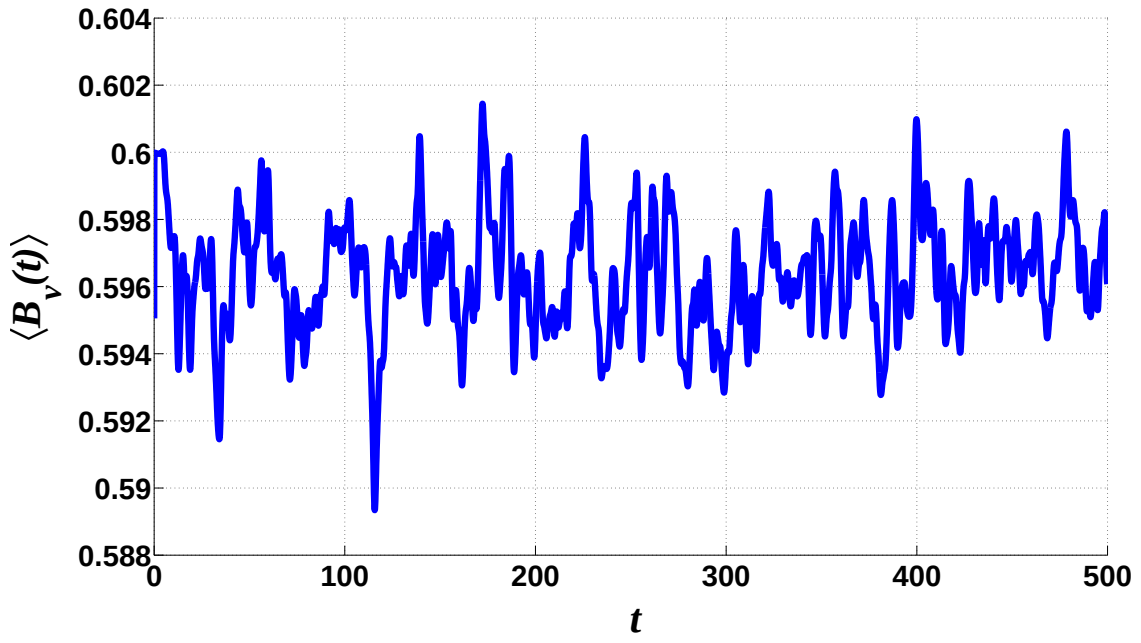

(b) $\langle B_v(t)\rangle$ for $\nu = 8$ exhibits chaos.

**Figure 8: Active cyber defense dynamics exhibit the chaos phenomenon: $G_B = G_R$ with $|V| = 2,000$ and $p = 0.005$.**

Figure 8(a) plots the MLE with respect to $\nu$. We observe that MLE $> 0$ when $\nu > 5$, meaning that system (17) exhibit chaos for $\nu > 5$. Figure 8(b) plots the phase portrait of $\langle B_v(t)\rangle$ (i.e., the average of the $B_v(t)$'s for all $v \in V$) when $\nu = 8$, which hints the emergence of chaos. This means that the defender should strive to avoid the parameter regime $\nu > 5$. This leads to the following:

INSIGHT 4. *Active cyber defense dynamics can be chaotic, meaning that it is impossible to predict the global cyber security state because it is too sensitive to the accuracy of the estimated initial global security state. Therefore, the defender must strive to avoid such unmanageable situations (e.g., by disrupting the attacks to assure $\nu \leq 5$ in the above example).*

## 7. RELATED WORK

Cybersecurity Dynamics is a framework for modeling and quantifying cyber security from a holistic perspective (rather than modeling and analyzing security of components or building-blocks) [34, 35, 36, 17]. This framework builds on a large body of literature across Computer Science, Mathematics and Statistical Physics (cf. [7, 4, 33, 37, 38, 17, 39, 29, 6, 3, 28, 23, 11, 12] and the references therein), which can be further traced back to the century-old studies on biological epidemic models [19, 13, 8].

As a specific kind of cybersecurity dynamics, active cyber defense dynamics were first *rigorously* modeled and studied in [36], despite that the idea of active defense has been discussed and debated for many years [14, 31, 18, 16, 26, 30, 1, 2]. We move a significant step beyond [36], by separating the *attack network structure* from the *defense network structure*, and by considering more general attack and defense power functions. To the best of our knowledge, we are the first to show that bifurcation and chaos are relevant in the cyber security domain, and to discuss the cyber security implications of these phenomena. Following [36], Lu et al. [17] investigate optimal active defense strategies in the Control-Theoretic and Game-Theoretic frameworks. Our study is complementary to [17] as we leave it to future work to investigate optimal strategies in our setting.

It is worth mentioning that models of Lotka-Volterra type [9] capture the predator-prey dynamics, which are however different from the active cyber defense dynamics. Active cyber defense dynamics may be seen as the *non-linear* generalization of the so-called Voter model in complex networks [25, 15]. Somewhat related to our work is [5], which considers chaotic dynamics in discrete-time limited imitation contagion model on random networks.

## 8. CONCLUSION

We have explored the rich phenomena that can be exhibited by active cyber defense dynamics. To the best of our knowledge, our study is the first to show that bifurcation and chaos are relevant in the cyber security domain. The implication is of high practical value: In order to make cyber security measurement and prediction feasible, the defender must manipulate the cyber security dynamics to avoid these *unmanageable situations*.

Interesting problems for future research include: First, we need to characterize non-homogeneous equilibria as we only focused on homogeneous equilibria. Second, we need to characterize which graph structure is more advantageous to the other (e.g., $G_B$ is ER graph but $G_R$ is power-law graph). Third, we need to explore the chaos phenomenon further (e.g., multi-direction chaos). Fourth, we need to systematically validate the models.

**Acknowledgement**. We thank the reviewers for their useful comments and Marcus Pendleton for proofreading the paper. Wenlian Lu was supported in part by the National Natural Sciences Foundation of China under Grant No. 61273309, the Program for New Century Excellent Talents in University (NCET-13-0139), the Programme of Introducing Talents of Discipline to Universities (B08018), and the Laboratory of Mathematics for Nonlinear Science, Fudan University. Shouhuai Xu was supported in part by ARO Grant #W911NF-12-1-0286 and NSF Grant #1111925. Any opinions, findings, and conclusions or recommendations expressed in this material are those of the author(s) and do not necessarily reflect the views of any of the funding agencies.

# APPENDIX

Now we prove Theorem 1.

PROOF. We prove the theorem in the first statement with $\mathbf{B}(0) \in \Xi_{G_B,\tau_1^*}$, and the second statement with $\mathbf{R}(0) \in \Xi_{G_R,\tau_2^*}$ can be proved similarly.

First, we see that $g(1) = 0$ implies that $B^* = 1$ is an equilibrium of (2) according to Proposition 1. Define

$$V_t = \operatorname*{argmin}_{v \in V} B_v(t) = \left\{ u \,\middle|\, B_u(t) = \min_{v \in V} B_v(t) \right\}$$

for $t \geq 0$. Since the case $\min_v B_v(0) = 1$, namely $B_v(t) = 1$ for all $v \in V$ and $t \geq 0$, is trivial, we assume $\min_v B_v(0) < 1$ without loss of any generality. For any $v(0) \in V_0$, the given condition (10) implies $\frac{1}{\deg(v(0),G_B)} \sum_{u \in N_{v(0),G_B}} B_u(0) \geq \tau_1^*$, and thus we have

$$\begin{aligned} & f\left(\frac{1}{\deg(v(0),G_B)} \sum_{u \in N_{v(0),G_B}} B_u(0)\right) \\ \geq \quad & \alpha \cdot \frac{1}{\deg(v(0),G_B)} \sum_{u \in N_{v(0),G_B}} B_u(0), \end{aligned}$$

where "=" holds only when $\frac{1}{\deg(v(0),G_B)}\sum_{u\in N_{v(0),G_B}} B_u(0) = 1$. Let $t=0$ and $v=v(0)$. Using Eq. (2) and condition (11), we have

$$\begin{aligned}
&\left.\frac{dB_{v(0)}(t)}{dt}\right|_{t=0} \\
= \;& f\left(\frac{1}{\deg(v(0),G_B)}\sum_{u\in N_{v(0),G_B}} B_u(0)\right)\Big[1-B_{v(0)}(0)\Big] - \\
& g\left(\frac{1}{\deg(v(0),G_B)}\sum_{u\in N_{v(0),G_B}} B_u(0)\right)B_{v(0)}(0) \\
\geq \;& f\left(\frac{1}{\deg(v(0),G_B)}\sum_{u\in N_{v(0),G_B}} B_u(0)\right) - \alpha B_{v(0)}(0) \\
\geq \;& \alpha\Big(B_{v(0)}(0) - B_{v(0)}(0)\Big) \qquad (17)\\
= \;& 0.
\end{aligned}$$

Since the equality signs hold in the two inequalities in Eq. (17) only when $\min_v B_v(0)=1$, which corresponds to the trivial case mentioned above, we conclude that $\min_{v\in V} B_v(t)$ strictly increases in a small time interval starting at $t=0$ except for the trivial case.

Let $\tau_1^{**} > \tau_1^*$ such that $\frac{1}{\deg(v(0),G_B)}\sum_{u\in N_{v(0),G_B}} B_u(0) > \tau_1^{**}$ for all $v\in V$. We now show that $\frac{1}{\deg(v,G_B)}\sum_{u\in N_{v,G_B}} B_u(t) > \tau_1^{**}$ for all $t>0$ and for all $v\in V$. Let $t_0$ be the first time that $\frac{1}{\deg(v,G_B)}\sum_{u\in N_{v,G_B}} B_u(t) = \tau_1^{**}$ for some $v\in V$, i.e.

$$t_0 = \inf\left\{\tau \,\middle|\, \frac{1}{\deg(v,G_B)}\sum_{u\in N_{v,G_B}} B_u(t) > \tau_1^{**} \forall t\in[0,\tau), \forall v\in V\right\}.$$

We show $t_0=+\infty$. Suppose $t_0<+\infty$. Let $V^*$ be the node set such that for each $v\in V^*$, $\frac{1}{\deg(v,G_B)}\sum_{u\in N_{v,G_B}} B_u(t)$ reaches $\tau_1^{**}$ for the first time. Then, for some $v^*\in V^*$, we know that $\frac{1}{\deg(v^*,G_B)}\sum_{u\in N_{v^*,G_B}} B_u(t)$ is not increasing at $t=t_0$. However, it can be shown that

$$\begin{aligned}
&\left.\frac{d}{dt}\left(\frac{1}{\deg(v^*,G_B)}\sum_{u\in N_{v^*,G_B}} B_u(t)\right)\right|_{t=t_0} \\
= \;& \frac{1}{\deg(v^*,G_B)}\sum_{u\in N_{v^*,G_B}} \left.\frac{dB_u(t)}{dt}\right|_{t=t_0} \\
\geq \;& \frac{\alpha}{\deg(v^*,G_B)} \cdot \\
& \sum_{u\in N_{v^*,G_B}}\left(\frac{1}{\deg(u,G_B)}\sum_{w\in N_{u,G_B}} B_w(t) - B_u(t_0)\right) \\
\geq \;& 0,
\end{aligned}$$

where the equality signs hold only for the trivial case as in the case of Eq. (17) mentioned above (i.e., in all other cases the inequalities are strict). So we reach a contradiction, which means $t_0=+\infty$. Owing to $\tau_1^{**}>\tau_1^*$, we have $\frac{1}{\deg(v,G_B)}\sum_{u\in N_{v,G_B}} B_u(t) > \tau_1^*$ for all $t>0$. That is, $\mathbf{B}(t)\in \Xi_{G_B,\tau_1^*}$ for all $t$.

Let $t_1$ be the maximum time that $\min_{v\in V} B_v(t)$ is strictly increasing, i.e

$$t_1 = \sup\left\{t \,\middle|\, \min_v B_v(t) \text{ is strictly increasing in } [0,t)\right\}.$$

We show that $t_1=+\infty$. Suppose that $t_1$ is finite, meaning that $\min_{v\in V} B_v(t)$ is not increasing at time $t=t_1$. Since it holds that $\min_{v\in V} B_v(t_1) > \min_{v\in V} B_v(0) > \tau_1^*$, by replacing $\mathbf{B}(0)$ with $\mathbf{B}(t_1)$, we have

$$\begin{aligned}
& f\left(\frac{1}{\deg(v(t_1),G_B)}\sum_{u\in N_{v(t_1),G_B}} B_u(t_1)\right) \\
> \;& \frac{\alpha}{\deg(v(t_1),G_B)}\sum_{u\in N_{v(t_1),G_B}} B_u(t_1)
\end{aligned}$$

and therefore we can show

$$\begin{aligned}
&\left.\frac{dB_{v(t_1)}(t)}{dt}\right|_{t=t_1} \\
\geq \;& f\left(\frac{1}{\deg(v(t_1),G_B)}\sum_{u\in N_{v(t_1),G_B}} B_u(t_1)\right) - \alpha B_{v(t_1)}(t_1) \\
\geq \;& \alpha\left(\frac{1}{\deg(v(t_1),G_B)}\sum_{u\in N_{v(t_1),G_B}} B_u(t_1) - B_{v(t_1)}(t_1)\right) \\
\geq \;& 0,
\end{aligned}$$

where are inequalities are strict except for the trivial case — as discussed in the case of Eq. (17). That is, $\min_{v\in V} B_v(t)$ strictly increases at $t=t_1$, which contradicts with the definition of $t_1$. Therefore, we have $t_1=+\infty$ and $\min_{v\in V} B_v(t)$ is strictly increasing in $t\in[0,+\infty)$.

In order to show $\lim_{t\to\infty} B_v(t)=1$ for all $v\in V$, we will prove that $\lim_{t\to\infty}\min_{v\in V} B_v(t)=1$ for $\lim_{t\to\infty}\min_{v\in V} B_v(t) \leq \lim_{t\to\infty} B_v(t)$. Since $B_v(t)$ is the probability that node $v\in V$ is blue at time $t$, we have $0\leq B_v(t)\leq 1$ for all $v\in V$. Hence $\lim_{t\to\infty}\min_{v\in V} B_v(t)$ exists. Suppose for the sake of contradiction that $\lim_{t\to\infty}\min_{v\in V} B_v(t)<1$, meaning $\min_{v\in V} B_v(t) < 1$ for all $t$ due to its strict increasing monotonicity. For any $v(t)\in V_t$, under the condition that Eq. (10) holds, there exists $\varepsilon>0$ such that $f(B_{v(t)}(t)) - \alpha B_{v(t)}(t) > \varepsilon$ for all $t$.

Since $\min_{v\in V} B_v(t)$ is strictly increasing for $t\in[0,+\infty)$, there exists $T>0$ such that

$$\begin{aligned}
&\frac{dB_{v(t)}(t)}{dt} \\
= \;& f\left(\frac{1}{\deg(v(t),G_B)}\sum_{u\in N_{v(t),G_B}} B_u(t)\right)\Big[1-B_{v(t)}(t)\Big] - \\
& g\left(\frac{1}{\deg(v(t),G_B)}\sum_{u\in N_{v(t),G_B}} B_u(t)\right)B_{v(t)}(t) \\
\geq \;& f\Big(B_{v(t)}(t)\Big) - \alpha B_{v(t)}(t) > \varepsilon,
\end{aligned}$$

for all $t>T$. This leads to

$$B_{v(t)}(t) > B_{v(T)}(T) + \varepsilon(t-T).$$

Since $\min_{v\in V} B_v(t) = B_{v(t)}(t)\to\infty$ as $t\to\infty$, it contradicts with $B_v(t)\leq 1$. Therefore, we conclude $\lim\limits_{t\to\infty}\min\limits_{v\in V} B_v(t)=1$ and $\lim\limits_{t\to\infty} B_v(t)=1$. $\square$